\documentclass[aps,prl,twocolumn,groupedaddress,showpacs]{revtex4}

\usepackage{graphicx}

\begin{document}

\title{ Doping and Irradiation Controlled Vortex Pinning in BaFe$_{2}$(As$_{1-x}$P$_{x}$)$_{2}$
 Crystals}

\author{ L. Fang$^{1}$, Y. Jia$^{1}$, J. A. Schlueter$^{1}$, A. Kayani$^{2}$, Z. L. Xiao$^{1}$, H. Claus$^{1}$, U. Welp$^{1}$, A. E. Koshelev$^{1}$, G. W. Crabtree$^{1}$, and W.-K. Kwok$^{1}$}
\affiliation{$^{1}$Materials Science Division, Argonne National
Laboratory, Argonne, IL 60439, USA } \affiliation{$^{2}$Physics
Department, Western Michigan University, Kalamazoo, MI 49008A, USA}
\begin{abstract}
We report on the systematic evolution of vortex pinning
behavior in isovalent doped single crystals of BaFe$_{2}$(As$_{1-x}$P$_{x}$)$_{2}$. Proceeding from optimal doped to ovedoped samples, we find a clear transformation of the magnetization hysteresis from a ‘fishtail’ behavior to a distinct peak effect followed by a reversible magnetization and Bean Livingston surface barriers. Strong point pinning dominates the vortex behavior at low fields whereas weak collective pinning determines the behavior at higher fields. In addition to doping effects, we show that particle irradiation by energetic protons can tune  vortex pinning in these materials.  
\end{abstract}


\maketitle

The discovery of multi-band superconductivity in iron pnictides \cite{JACS} with
relatively high transition temperatures and modest superconducting
anisotropy has opened new research approaches for realizing an isotropic
high $T_{C}$ superconductor.  These materials share as common structural motif superconducting FeAs or FeSe/Te layers.  Generally, superconductivity emerges from a semi-metallic antiferromagnetic parent compound \cite{review}, upon electron, hole or isovalent doping, BaFe$_{2}$(As$_{1-x}$P$_{x}$)$_{2}$ being an example of the latter \cite{BaFeAsPtrans}. A key feature that can strongly affect the behavior of multi-band superconductors is the interplay of inter and intra-band electron scattering.  The scattering mechanism does not only affect the superconducting gap structure \cite{interbandscattering}, the temperature dependence of thermodynamic quantities such as the upper critical field \cite{interbandHc2} and the superfluid density \cite{interbandsuperfluid}, but also the pinning of superconducting vortices \cite{interbandvoties}.  The charged dopant sites in electron and hole doped materials create strong scattering potentials that dominate electron scattering.  In contrast, in isovalent doped BaFe$_{2}$(As$_{1-x}$P$_{x}$)$_{2}$ (BFAP)  scattering can be tuned from the clean limit to the dirty limit; in pristine strongly-over-doped BFAP, de Haas van Alphen experiments have revealed long electron mean free paths \cite{quantumoscillation1,quantumoscillation2}.

\begin{figure}
\includegraphics[width=9cm]{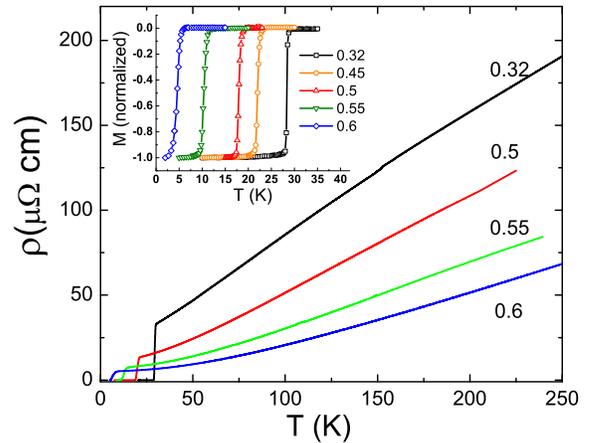}
\caption{(Color online) Temperature dependence of the resistivity in zero field for BFAP.  (inset) Superconducting diamagnetic transition of BaFe$_{2}$(As$_{1-x}$P$_{x}$)$_{2}$ crystals (0.32 $\leq x \leq$ 0.6) measured at $H$ // $c$ = 10 Oe.   } \label{fig1}
\end{figure}

In this work, we report on a systematic study of vortex pinning in a series of doped BFAP crystals. We find a pronounced magnetization ‘fishtail’ behaviour, which evolves into a peak effect (PE) in the critical current with increased phosphorus doping.  With further doping, vortex pinning becomes virtually negligible, leaving only the Bean Livingston (BL) barrier as a source of vortex pinning, underlining the purity of the sample with high phosphorus content. We demonstrate that strong point pinning (SPP) dominates at low magnetic fields, whereas weak collective pinning (WCP) occurs at high fields.  In addition to doping effects, we show that irradiation by protons can tune the vortex pinning behavior in these materials.

 High
purity BaFe$_{2}$(As$_{1-x}$P$_{x}$)$_{2}$ (x=0.32$\sim$0.6) single
crystals were grown using a self flux method. The elemental
composition was determined using $X$-ray Energy Dispersion Spectra. 
The crystals were cut into
a rectangular shape with size approximately 400 x 300 $\mu m$$^{2}$ for
magnetization and resistivity measurements. The upper critical field, $H_{C2}$, was determined from linear extrapolations of the temperature dependent reversible magnetization to the normal state baseline. The irreversibility field, $H_{irr}$, was
determined from the first separation point between the zero field
cooled and field cooled magnetization curves. The critical current
density ($j_{C}$) was estimated using the Bean critical state model. In the case of dominant surface pinning, supercurrents are concentrated at the sample surface. We nevertheless calculated $j_{C}$ values for comparison of the field dependence among samples. One over-doped crystal (x=0.55) was irradiated at the tandem
accelerator at Western Michigan University with 2 MeV protons to a
dose of 8$\times$10$^{15}$ p/cm$^{2}$.

\begin{figure*}
\includegraphics[width=17cm]{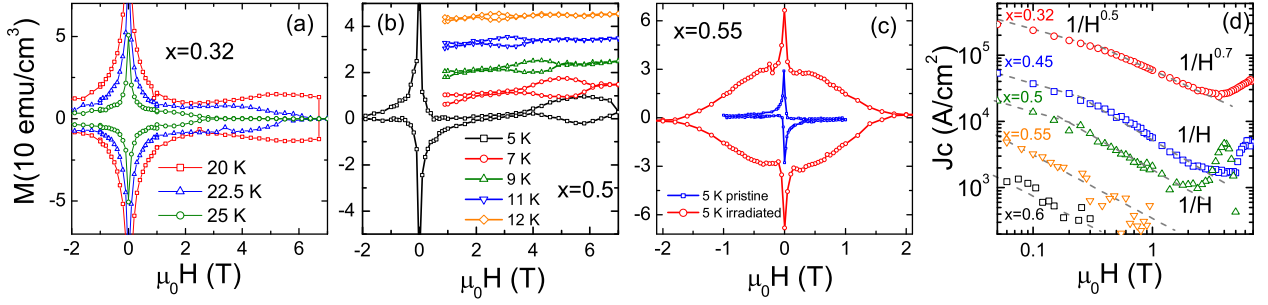}
\caption{(Color online)Magnetization hysteresis of (a) optimum doped (x=0.32) BFAP, (b) overdoped (x=0.5) (note, curves for different temperatures shifted for clarity) and (c) pristine and proton irradiated overdoped (x=0.55) crystal for $H$ // $c$.  (d) $j_{C}(H)$ for optimal and overdoped crystals measured at the same reduced temperature $T/T_{C} \cong 0.55$. The dashed lines are guides to the eye. } \label{fig2}
\end{figure*}

The inset of Fig. 1 shows the temperature dependent magnetization curves  for doping
from x=0.32 to 0.6. Except for 
x=0.6, the transition widths of the crystals are very sharp with
$\Delta T_{C}$ (10\% - 90\%)$<$ 1 K. The corresponding resistivity curves are shown in the main panel. The residual resistivity decreases strongly upon P-doping from 30 to less than 5 $\mu$$\Omega$ cm underlining the increased purity of the over-doped samples.   Fig. 2a displays the field
dependent magnetization for the optimal doping (x=0.32) at various
temperatures. The magnetization shows a sharp peak centred at zero field as observed in most studies on the FeAs-superconductors and, in contrast to a previous study\cite{strongpinning},  a broad
maximum indicative of the 
'fishtail' effect at high fields.  The fishtail peak moves to lower
fields with increasing temperature.  The field and temperature dependence of the fishtail is
similar to that observed in other pnictides \cite{NdFeAsO,BaFeCoAs,BaKFeAs,BaKCoAsHHWen,BaKFeAsBrazil,BaFeCoAsAPL}.
With increasing phosphorus doping (x= 0.5), the 
magnetization hysteresis curve collapses and becomes asymmetric, and nearly reversible at intermediate fields, followed by a pronounced PE  at higher fields as shown in Fig. 2b.  
 The
PE moves to lower fields and its magnitude diminishes with
increasing temperature. No PE was observed when
the field was applied parallel to the crystal's ab-plane.
With further phosphorus doping, x=0.55, the PE disappears and the magnetization hysteresis curve at T = 5 K becomes almost reversible over its entire field range up to the upper critical field.  Furthermore, the descending field branch of the magnetization curve shows nearly zero magnetization. Such behaviour has been observed in very clean YBa$_{2}$Cu$_{3}$O$_{6+\delta}$ crystals and was considered as a fingerprint of vortex pinning by the BL surface barrier\cite{ybcosurfacebarrier, BLbarrier},which becomes the dominant pinning mechanism at high fields when vortices are straight in the absence of bulk pinning.

\begin{figure}
\includegraphics[width=8cm] {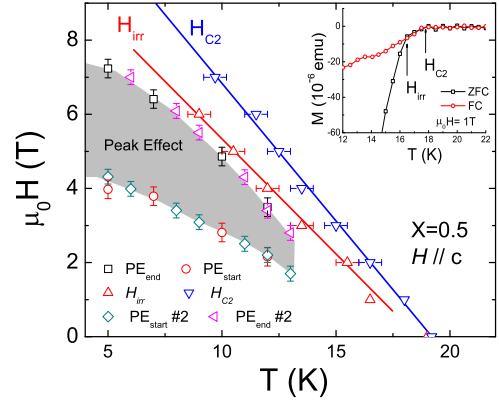}
\caption{(Color online) Phase diagram of the x=0.5 sample. Inset shows the determination of the upper critical field and the irreversibility field through field cooling and zero field cooled temperature dependent magnetization curves. } \label{fig3}
\end{figure}

Fig. 2d compares the critical currents, $j_{C}$($H$), of all the phosphorus doped crystals obtained at the $\textit{same reduced temperature}$, $T$/$T_{C}$ $\approx$ 0.55. A
common feature displayed in Fig. 2d is the $H^{-0.5}$
dependence at low fields below $H <$ 0.3 T and the $H^{-0.7}$ to
$H^{-1}$ dependencies at higher fields for the 0.32 $<$ x $<$ 0.5
samples. These $j_{C}$($H$) behaviors  are consistent with SPP theory \cite{strongpinning theory}. In the simplest
case when all pins are occupied by vortices,
$j_{C}=\frac{n_{p}f_{pin}}{B}$, where $n_{p}$ is the volume
concentration of effective pinning centers, $f_{pin}$ is the maximum
value of pinning force on one defect and $B$ is the magnetic
induction.  At smaller fields when interaction between vortices
prevents occupation of all pin sites,
$F=\Phi_{0}j_{C}=\varepsilon_{0}n_{p}b\sqrt{\frac{\Phi_{0}}{B}}$ ,
where $\Phi_{0}$ is the flux quantum, $b$ is the
size of the pinning defect and assumed greater than coherence $\xi$,
$\varepsilon_{0}=\Phi_{0}^{2}/4\pi\mu_{0}\lambda_{ab}^{2}$ is the
characteristic vortex energy per unit length, $\mu_{0}$ is the
magnetic constant and $\lambda$ is the penetration depth.  Hence
both $H^{-1}$ and $H^{-0.5}$ relations can be accounted by SPP.
Similar SPP behavior has recently been observed in several 122 and 1111-based
pnictides at fields below the  appearance of fishtail behavior
\cite{strongpinning, NdFeAsO}. In elevated fields, typically larger than 1 T, SPP becomes negligible and is overtaken by collective pinning as evidenced by the observation of a fishtail at optimal doping and a PE with at higher phosphorus concentration (see Fig. 2d).  Fishtail and PE are typically associated with a large concentration of weak pinning sites, leading to collective pinning behaviour.

A $1/H$ dependence of $J_{C}$ is also observed for the highly over-doped samples, albeit at very low $J_{C}$ values. This behavior is expected for BL barrier pinning \cite{BLJc}, where $m\approx H_{p}^{2}/2B$ with $m$ the magnetic moment on the ascending branch, and $H_{p}$ is the first field of flux penetration. Thus, considering the shape of the magnetic hysteresis of the x=0.55 sample, the $1/H$ dependence is associated with the BL barrier instead of SPP. Fig. 2d also shows that $J_{C}$ drops strongly from$ \sim 4\times 10^{5} A/cm^{2}$ to $\sim 2\times 10^{4} A/cm^{2}$ (at reduced temperatures of $T/T_{C}$ = 0.55) in going from x= 0.32 to x=0.5 when bulk pinning is dominant.   The $J_{C}$-value for x = 0.32 is in good agreement with a previous report\cite{BFAPcriticalcurrent}.  The doping dependence of $J_{C}$ is also consistent with recent de Haas van Alphen  measurements in overdoped BFAP that reported an increase in the mean free path with  phosphorus doping \cite{quantumoscillation1,quantumoscillation2} . However, it differs from a report in which a higher $J_{C}$ was found for a x=0.49 sample than for x=0.32 at $T/T_{C}$ = 0.3\cite{strongpinning}.

To test whether irradiation induced defects could mimic the effect of vortex behaviour due to phosphorus doping, we  irradiated the pristine x=0.55 crystal which showed essentially no bulk pinning (see Fig. 2c).   Fig. 2c shows the magnetic hysteresis curve obtained at 5 K on the same sample after irradiation . The magnetization hysteresis is enhanced considerably compared to that of the pristine crystal indicating that the p-irradiation induced defects are effective pinning sites.  The $M(H)$ curve does not display a discernible fishtail, and the extent of the enhanced pinning at low fields is strongly reduced as compared with the data in Fig. 2a, for instance.

\begin{figure}
\includegraphics[width=7cm]{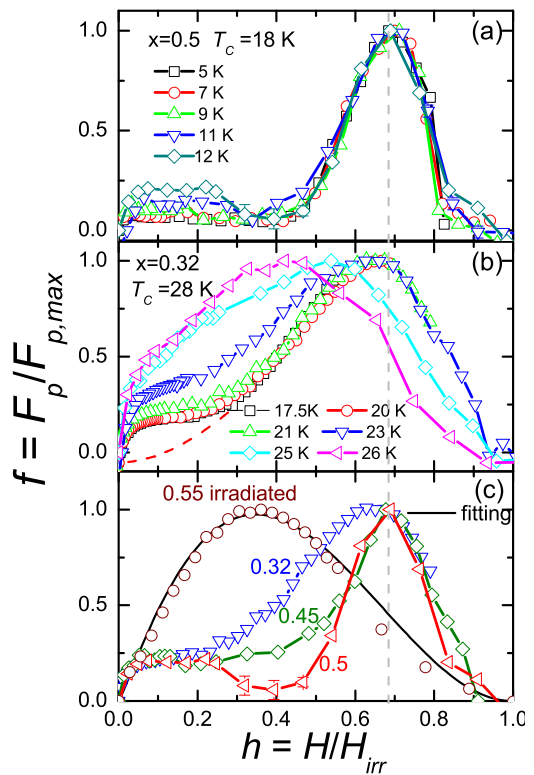}
\caption{(Color online)(a) Normalized pinning force as a function of reduced field for BFAP with x=0.5 at various temperatures, and (b) with x=0.32. The dotted red line is a guide to the eye indicating the contribution due to collective pinning to the pinning force. (c) Comparison of $f(h)$ among various phosphorus doped samples and a proton irradiated x=0.55 sample at $T/T_{C}$ = 0.66} \label{fig4}
\end{figure}

Fig. 3 shows the phase diagram as deduced from temperature dependent magnetization measurements.  We determine the slope of the upper critical field of $\mu_{ 0}dH_{C2}/dT$ = -0.86 T/K, in good agreement with results of our specific heat measurements.  The phase diagram is characterized by a narrow vortex liquid region  between $H_{C2}$ and $H_{irr}$, and by the occurrence of the peak effect just below $H_{irr}$.  These characteristics strongly resemble those of MgB$_{2}$ \cite{MgB2phasediagram1,MgB2phasediagram2}.  However, on our current samples  the resistivity nor the magnetization displays a discontinuous temperature dependence that could be indicative of a first order vortex transition.

Plots of the normalized pinning force ($f = F_{p}/F_{p, max} , F_{p} = \mu_{0} H \times j_{C}$ ) as a function of reduced field ($h = H/H_{irr}$) have proven useful for identifying various pinning regimes\cite{BaKFeAs, BaFeCoAsAPL, Dewhuges}.  The pinning force curves for x = 0.5 (Fig. 4a) are characterized by a sharp maximum near $h$ = 0.7 (corresponding to the PE) and a plateau at $h$ $<$ 0.2 corresponding to the approximate $1/H$-dependence of $j_{C}$ and thus representing the regime of SPP.  The curves in the PE-region display remarkable scaling for different temperatures.  At temperatures above 12 K the PE disappears and the magnetization in high fields becomes reversible. For the x = 0.32 sample we observe a similar, albeit broader, maximum around $h$ = 0.7, corresponding to the fishtail in the magnetization.  The plateau at $h$ $<$ 0.2 as well as the steep rise at very low fields represent the strong point pinning regime, since the initial $j_{C} \sim H^{-1/2}$ dependence would correspond to a square-root rise in the pinning force. The pinning force curves scale well at temperatures below 23 K, however this scaling breaks down at the highest temperatures and the $f(h)$-curves acquire a more symmetric shape. Fig. 4c displays the progression of the $f(h)$ for various doping levels at  $T/T_{C}$ = 0.66.  These results indicate that starting from a certain degree of bulk pinning, here achieved in the sample with x = 0.5 at temperatures below 13 K, a sharp peak effect appears near the irreversibility line.  With increased bulk pinning this peak effect widens into the fishtail feature in the magnetization hysteresis.  It is a remarkable feature of BFAP that over large sections of the $H-T-x$ diagram, pinning can be separated into two distinct mechanisms: strong point pinning at low fields and collective pinning at high fields.  Even though critical currents due to different pinning mechanisms are not simply additive in general, the data in Fig. 4b  suggest that the pinning force due to the collective pinning mechanism is negligible at low fields. 

Also included in Fig. 4c is the pinning force curve of the p-irradiated x = 0.55 sample.  It has a conventional shape with no signs of strong point pinning at low fields, and is fitted well by the relation $h(1-h)^{2}$.  This type of pinning force curve is expected  for pinning by normal point pins ($\delta T_{C}$ pinning).  In general, however the pinning force curves in Fig. 4 do not follow the standard forms\cite{Dewhuges}. It is nevertheless instructive to note that a maximum of the pinning force near $h \sim $0.7 is indicative for $ \delta \kappa$ ($\delta \textit{l}$) pinning.  This pinning mechanism has been invoked in the previous analysis \cite{NdFeAsO, BFAPcriticalcurrent}  of the pinning properties of 1111- and 122- based samples.  It is likely to arise due to fluctuations in the electron mean free path caused by inhomogeneity in the dopant distribution.  As the sample temperature approaches $T_{C}$ variations in $T_{C}$ become relatively more important leading to areas of suppressed condensation energy or normal regions.  In this case $\delta T_{C}$-pinning is expected, characterized by a maximum in the pinning force curve near 0.35.  This scenario can account for the temperature evolution of the pinning force curves of the x = 0.32 sample, and the results for the p-irradiated sample if one assumes that the irradiation induced defects are regions of suppressed superconductivity\cite{protrondefectsize}.  Upon increasing temperature, the x = 0.5 (and 0.45) sample transitions into the state of the x = 0.55 sample, that is, vanishing bulk pinning before the regime of $\delta T_{C}$-pinning can be reached.

These general features are not limited to BFAP but arise in materials that can be synthesized with fairly high purity, such as MgB$_{2}$. Neutron as well as electron irradiation studies on MgB$_{2}$ crystals have also shown the transformation from a virtually reversible magnetization curve in clean MgB$_{2}$ crystals into a PE and eventually into a fishtail shaped magnetization \cite{MgB2phasediagram2, neutron2} . Our data clearly demonstrates that the PE emerges from the fishtail magnetization, once  pinning is reduced via phosphorus doping.  Conversely, the large magnetization hysteresis can be recovered by introducing point (or cluster) defects via proton irradiation.  Our studies also point to the possibility of separating the effects of strong and weak collective pinning via doping and controlled proton irradiation.

Acknowledgement: This work was supported by the Center for Emergent
Superconductivity, an Energy Frontier Research Center funded by the
U.S. Department of Energy, Office of Science, Office of Basic Energy
Sciences under Award Number DE-AC0298CH1088 (LF, YJ, HC, AEK, GWC,
WKK) and by the Department of Energy, Office of Basic Energy
Sciences, under Contract No. DE-AC02-06CH11357 (JAS, ZLX, UW).

\end{document}